# *The Physics of Mind and Thought*

# Brian D. Josephson

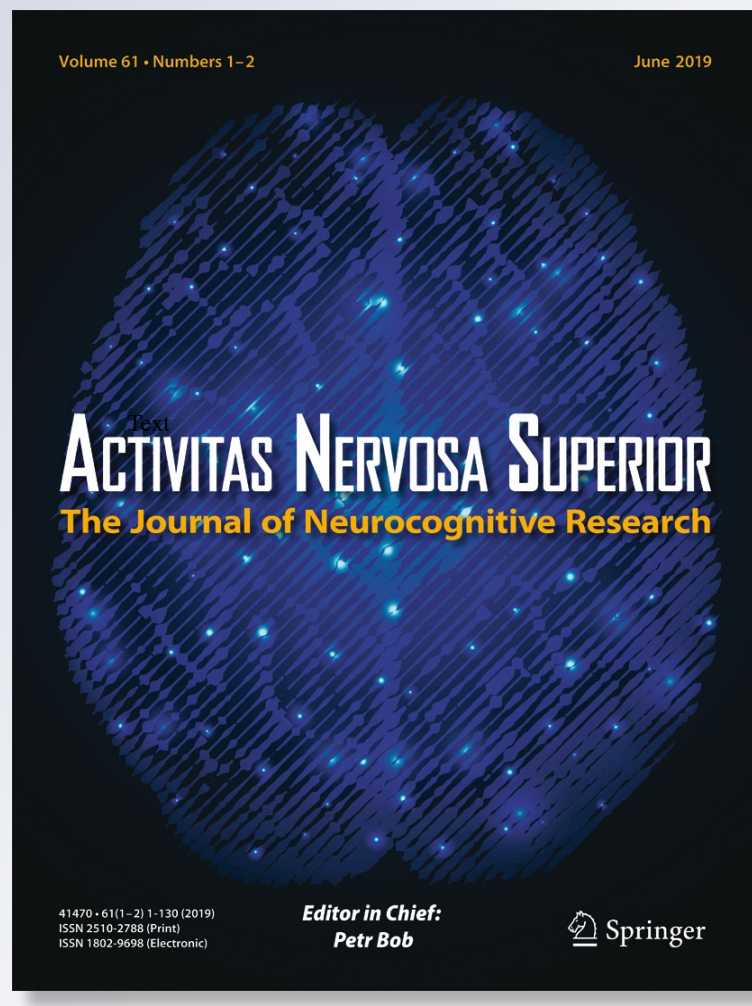

Springer

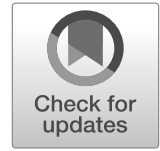

# The Physics of Mind and Thought

Brian D. Josephson[1]

## Abstract
Regular physics is unsatisfactory in that it fails to take into consideration phenomena relating to mind and meaning, whereas on the other side of the cultural divide such constructs have been studied in detail. This paper discusses a possible synthesis of the two perspectives. Crucial is the way systems realising mental function can develop step by step on the basis of the scaffolding mechanisms of Hoffmeyer, in a way that can be clarified by consideration of the phenomenon of language. Taking into account such constructs, aspects of which are apparent even with simple systems such as acoustically excited water, as with cymatics, potentially opens up a window into a world of mentality excluded from conventional physics as a result of the primary focus of the latter on the matter-like aspect of reality.

## Introduction: the Problem of Mind

Henry Stapp (Stapp 2016) is one of a number of scientists who have felt quantum mechanics to be an incomplete account of nature, failing to give a satisfactory account of observation, in which context it appears that an aspect of mind, namely the intention and attention of the observer, is a factor relevant to the outcome. Wheeler (1983) took this concept further, proposing that over time quantum observation, carried out in an appropriate manner, might lead to what he referred to as the 'fabrication of form'. He did not however offer a detailed analysis. The whole question of quantum observation is a confused one, suggesting the need for a new approach, based on some kind of unification involving phenomena of mind, not covered by current approaches, on the one hand, and matter, which is so covered, on the other.

Similar situations have been encountered before in physics. For example, at one time it was assumed that transmutation of elements was impossible, but then radioactivity was discovered and, as a result, a whole new landscape became accessible to physics. In the case of mind, while it is accepted that phenomena involving mind, such as problem solving, do occur, such phenomena are generally considered irrelevant to physics, with the result that a similar extension in the scope of physics has not occurred. There is a kind of 'two cultures' divide, where people who are interested in the mechanisms of mind are generally unconcerned with its possible relevance to physics. There have been exceptions, such as Bohm (1985), but the writings of such people are typically dismissed by the mainstream.

The fact that mentality, meaning productive information processing, is something very different to anything regular physics has to offer can be clearly illustrated by considering questions such as 'what is a lecture', which cannot be addressed at all within the conventional frameworks. From this, it would seem that words have, in some way, meaning, which meaning is extracted, and used fruitfully, by associated structures. Bohm suggested, with his concept of soma-significance, that matter may be associated with indefinite depths of meaning while again Barad (2007) discussed, in connection with a picture involving a concept known as agential realism, 'the entanglement of matter and meaning'. In the following, a somewhat different approach is taken, perhaps more appropriate to the task of integration into physics, that hopefully will have the capacity to transform radically our understanding of nature in the same way that our understanding of the details of biology transformed our understanding of life. This approach involves a synthesis of a number of existing approaches.

✉ Brian D. Josephson
bdj10@cam.ac.uk

1 Cavendish Laboratory, 19 J J Thomson Avenue, Cambridge CB3 0HE, UK

## Signs, Semiotic Scaffolding, and Stepping Stones

On the other side of the cultural divide referred to above, there is the fundamental idea of a sign, the study of which originated in the semiotic concepts developed in the nineteenth century by Charles Sanders Peirce, more recently taken up by biologists, thereby founding the subject of *biosemiotics* (Hoffmeyer 2008a, 2008b). Semiotics emphasises the role of interpretation, a process connecting signs with corresponding objects, or more generally *mediation*, a process involving situations where a third entity influences the relationship between two others. Such mechanisms play a key role in biology, and one that is essential for effective biological function.

This points to something missing from the physical scientists' approach to biology (cf. Josephson 1988), namely that biology involves not only component systems but also the interrelationships between the component systems, which are not fixed in advance, but themselves develop through the mediation of a range of mechanisms. Thus, Hoffmeyer cites the example of the movement of an organism in response to a chemical stimulus (chemotaxis), which behaviour depends on 'a sophisticated interaction of some fifty different proteins'. He introduces, in this connection, the concept of the *semiotic scaffolding* for a biological process, applied to whatever influence it is that acts upon such a system for it to be able to produce the orderly behaviour necessary for the successful execution of the given process.[1] Such an influence involves the processing of information relevant to that process (and accordingly meaningful in that context), leading to the outputting of the code needed to select the action appropriate at the given time. Computer routines and biological systems both work in this way, with biological systems using chemical reactions or neural mechanisms to perform the required computations.

Hoffmeyer also introduces *stepping stones* as a metaphor for describing how systems develop, a stone being related to the knowledge a system has at a particular stage of development of some skill (equivalently, to a particular domain within a space of actions involving that skill). The advance in knowledge associated with stepping from one stone to an adjacent one is supported by scaffolding relevant to advancing in that specific direction. To take the metaphor further, some stones are less reliable than others, so that over time a steadily increasing collection of reliable stones emerges. These concepts collectively provide a picture of how skills advance, which will now be related to a different metaphor, that of a game.

---

[1] Subsequent to the submission of this paper, the author became aware of an approach to understanding the complexities of biological systems known as coordination dynamics (Kelso 2013), based upon study of the way component parts of unified entities coordinate with each other through the exchange of meaningful information. Such analyses are likely to clarify the scaffolding concept considerably, leading to more detailed insight into mechanisms of mind.

## Knowing the Game

We return now to the question, what is a lecture? According to Foucault (Gutting and Oksala 2018), such questions need to be addressed in historical or archaeological terms (and biosemiotics similarly notes the relevance of history to the understanding of the behaviour of a developing system). Thus, a lecture involves a particular set of practices that originated in the past, continuing because people 'know the game', an idea implicit in Wittgenstein's 'language games', and in the words of the ABBA song:

> we know the start, we know the end
> Masters of the scene
> We've done it all before
> and now we're back to get some more

Thus, participants in games such as these, with lectures as an example of such a 'game', follow procedures previously acquired, resulting in the achievement of some specific end.

A number of points can be made in this connection. In the case of language generally, a successful computer model involving a collection of specified procedures has been created by Winograd (1972), simply by realising in computer software theories as to what is involved in linguistic activity. Its success, exemplified by its ability to respond correctly, in the context of a simulated blocks world, to complicated input such as the question 'is there anything which is bigger than every pyramid but is not as wide as the thing that supports it?' supports a view of the structure of mind proposed here, and related to Minsky's 'society of mind' (Minsky 1986), that the diverse capacities of mental activity result simply from the workings of a hierarchy of specialised systems, each related to a range of needs arising in connection with the process as a whole consisting in effective communication.

In the case of Winograd's program, the details of the procedures implementing the specialisations concerned were put in by hand, but linguistic activity generally can be interpreted in terms of the stepping stone model, hypothesising the progressive incorporation into the system of such new possibilities as happen to be effective in particular situations. Such incorporation can be expected to be assisted by scaffolding embodying specific concepts, for example the idea that words may relate to objects or relationships in the environment. To this particular principle there can be expected to correspond a scaffolding mechanism that will look for such relationships, and reconfigure itself so as to respond appropriately in the future. Similar scaffolding mechanisms can be expected to exist regarding other aspects of language, their joint activities serving to develop particular languages over time, as well as leading to the acquisition of languages by individuals.

Communication through language exemplifies another general aspect of mind, the 'oppositional dynamics' of Yardley ([2010](#)), a reference to the strongly constraining influence of the need for systems to be able to work together, a necessity familiar in the way web servers and web browsers need to follow a specific protocol in order for information transfer to be effective. Pairs of systems in such a situation, such as those of speaker and listener in the case of language, or specification in code and what is specified by that code, need to establish how to achieve coordination, in general with the support of appropriate scaffolding. Once techniques relevant to a skill such as language have been established by a group of people, others can copy these techniques and take advantage of them.

## Abstractions, Symbolic Processes, and Truth

Sign use, as originally argued by Peirce and subsequently developed by Deacon ([1998](#)) and by Favareau ([2015](#)), is of three types, iconic, indexical, and symbolic. The first two types of sign involve entities in the immediate environment, but symbolic use, which use appears to be confined to human beings, can involve manipulations concerning entities absent from the immediate environment, which faculty is attributed by Deacon to a human ability to avoid being too involved in the current situation during mental activity. This can be understood in terms of memory mechanisms that can, as it were, hold on to signs so as to be able to act systematically with them, and thereby develop 'games' such as mathematics. Thus, it can be argued (Josephson [2012](#)) that geometry originated on the basis of activities involving physical lines and points, facts about which were idealised by Euclid, whose thinking led to a set of axioms and to derivative theorems, as a result of the step by step advances in competence of the kind discussed. The fact that signs, though originating in concrete reality, can subsequently be manipulated in ways not connected with the situation that an individual is in at the time makes possible 'symbolic fantasies'. This raises the question of what constrains such abstractions to make them meaningful. In the earlier indexical use of signs, involving relationships, *consistency* (related to oppositional dynamics) between symbolic activity and what is present in the environment is important, and this may apply equally to symbolic abstraction: for example, a mathematician may fantasise that minus one has a square root and then try to erect a consistent scheme at the level of symbolic activity, thereby creating a synthetic reality that others can act in. This is what conversation is about; individuals develop tools for creating a synthetic reality on the basis of their past experience (compare this with building real objects with a construction kit, on the basis of descriptions in language) and can cooperate in their use.

## Which Physics?

The discussion so far has largely featured an information-processing perspective, ignoring the question of the underlying physics. Its motivation was the idea (cf. Josephson [2002](#); Barad [2007](#)) that mind processes are relevant at levels other than familiar ones such as brains and are involved in contexts such as quantum observation. Thus, the question arises of what circumstances might give rise to mind-like phenomena. The little-known subject of *cymatics* (Reid [2017](#)) is of interest in that it shows that phenomena related to those of mind can occur even in apparently simple systems such as water. The field of study involves the use of an instrument known as the cymascope, used to observe the surface of water being excited acoustically with a signal that may for example be one at a specific frequency (Sheldrake and Sheldrake [2017](#)), or one derived from music (Buchanan [2012](#)). As the amplitude of the signal is increased, at some point an instability occurs, accompanied initially by chaos but, under certain conditions, the system settles down to a specific spatial pattern dependent upon the signal, with a superimposed oscillation. We have here in effect a system 'discovering' a stable response to a signal. More surprisingly, applying to the water a signal derived from an echo-locating dolphin is found to regenerate the shape of the object that was the source of the echo, which ability may depend on the formation of structure in the water. It remains to be seen exactly what underlies such phenomena, and whether higher levels of complexity, involving mind-like behaviour, can originate by similar means.

In the above, we have situations where two physical entities (e.g. oscillation of the water and its spatial pattern) each influence the other with, in some cases, a stable situation ultimately developing, while in other cases the systems remain effectively uncorrelated. This would seem to be an important physical mechanism underlying semiotic behaviour, a third system providing the background mechanism underlying the mutual influence while, in cases such as discussed in connection with language, units may come to combine in more and more complex ways, with configurations emerging possessed with overall stability, 'stepping' from one configuration to another as new potentially stable combinations emerge through chance encounters.

## Links with Quantum Mechanics

The question remains to what extent a unified approach, combining the above analyses with those of conventional physics, may be possible, involving for example the discovery of parallels between quantum phenomena and mind such as those discussed in Josephson ([2002](#)). Ideas as to how the two domains are related have also been discussed by Barad ([2007](#)), who for example discusses the 'apparatus' referred to by Bohr in connection with quantum observation in ways paralleling the semiotic scaffolding of Hoffmeyer, both of which have the effect of directing development in specific ways. Barad introduces also a concept of 'intra-actions' (interactions within a unitary system) that parallels both entanglement and oppositional dynamics in the way they involve two or more systems acting as one.

Another potential connection is one with 'reverse causation', a concept presuming an influence backwards in time (e.g. Aharonov and Gruss 2005), though in the current approach the backwards in time influence is merely apparent. Sutherland (2016) for example argues that specifying a state in terms of some future situation as well as the past enables one to restore locality in the case of entangled states and to replace statistical descriptions by ontological ones. But, to a large extent, the same applies in situations governed by mind-like processes, in which context reverse causation need not be assumed. To see this, consider the fact that if we know where a person is at the present time, then many factors need to be taken into account in predicting the future, and the conclusion may have to be expressed in statistical terms. If however we know where a person is going to be some time in the future, then the uncertainty may be removed, making it possible to predict the intermediate activity.

## Summary

The basic problem with quantum mechanics is that a person's decision as to what aspect of nature to observe can have real consequences, and it is unclear how such mental activity can be integrated with traditional physics; we cannot simply leave out the observer. A thesis in the above has been that semiotics (sign theory) will play a central role in such a future integrated physics, a basic task for such a future physics being that of bridging the gap between signs and the phenomena addressed by current physics, thereby arriving at an integrated point of view. A similar situation arises in conventional science, where a gap of this kind exists between fundamental physics and biology, one that can be bridged taking due account of a succession of levels, utilising a range of specialised approaches to deal with these.

In the present case, the basic concept is that of the semiotic triad, where one entity influences the relationships between two others. Under suitable circumstances, such relationships emerge spontaneously. From such a starting point, functional complexity can build up as has been discussed in connection with language. Further, language can evolve so as to be able to symbolise abstractions including mathematics, potentially leading, in line with Wheeler's 'fabrication of form', to the origination of universes subject to mathematical laws, and the possible influence of mind mechanisms on life processes in such a universe, and on the course of evolution. In principle, it should be possible for a detailed picture along such lines to be built up over the course of time, following traditional methodologies for evaluating ideas in physics such as that of constructing and testing suitable models. This will not be a straightforward process, but there may be no other way to advance beyond the unavoidable limitations associated with the outdated idea that the complexities of reality can be reduced to a formula.[2]

## Compliance with Ethical Standards

---

[2] Note that, for practical reasons, not all relevant research areas could be discussed in the above analysis, examples of such additional research being the neurosciences and neural networks, complexity biology (Hankey 2015), Kastner's version of the transactional interpretation of quantum mechanics involving 'the reality of possibility' (Kastner 2012) which, as in the present formulation, does not invoke reverse causation, and finally meaning associated with musical forms (Josephson and Carpenter 1996).

**Author's Note** (N.B.: additional note for arXiv version of the paper) Encyclopaedia Britannica defines philosophy of physics as 'philosophical speculation about the concepts, methods, and theories of the physical sciences', a category to which this paper arguably belongs. The physics preprint archive arXiv, however, refuses to allow it to be listed in its most relevant section, History and Philosophy of Physics (physics.hist-ph).